\pdfoutput=1

\documentclass[11pt,twoside,a4paper,cmspaper,final,collab]{cms-tdr}
\def\svnVersion{471738P}\def\cmsCernNoTag{CERN-EP-2018-109}\def\cmsCernDate{\today}\def\cmsMessage{Published in Physical Review D as \href{http://dx.doi.org/10.1103/PhysRevD.98.032005}{\doi{10.1103/PhysRevD.98.032005.}}}

\begin{document}\cmsNoteHeader{SUS-18-001}

\hyphenation{had-ron-i-za-tion}
\hyphenation{cal-or-i-me-ter}
\hyphenation{de-vices}
\RCS$HeadURL: svn+ssh://svn.cern.ch/reps/tdr2/papers/SUS-18-001/trunk/SUS-18-001.tex $
\RCS$Id: SUS-18-001.tex 467119 2018-07-02 18:19:13Z alverson $

\newlength\cmsTabSkip\setlength{\cmsTabSkip}{1ex}
\ifthenelse{\boolean{cms@external}}{\providecommand{\NA}{\ensuremath{\cdots}\xspace}}{\providecommand{\NA}{\ensuremath{\text{---}}\xspace}}
\providecommand{\cmsTable}[1]{\resizebox{\textwidth}{!}{#1}}

\newcommand{\fullLumi}{35.9\fbinv}
\newcommand{\mttwo}{\ensuremath{M_{\mathrm{T2}}}\xspace}
\newcommand{\vMht}{\ensuremath{\vec{H}_{\mathrm{T}}^{\text{miss}}}\xspace}
\newcommand{\njets}{\ensuremath{N_{\mathrm{j}}}\xspace}
\newcommand{\nbtags}{\ensuremath{N_{\PQb}}\xspace}
\newcommand{\njetsisr}{\ensuremath{N_\mathrm{j}^\mathrm{ISR}}\xspace}
\newcommand{\vjets}{\ensuremath{\cmsSymbolFace{V}\text{+jets}~(\cmsSymbolFace{V}=\PZ,\PW)}\xspace}
\newcommand{\wjets}{\ensuremath{\PW\text{+jets}}\xspace}
\newcommand{\zjets}{\ensuremath{\PZ\text{+jets}}\xspace}
\newcommand{\ttjets}{\ensuremath{\ttbar \text{+jets}}\xspace}
\newcommand{\znunu}{\ensuremath{\PZ\to\PGn\PAGn}\xspace}
\newcommand{\zll}{\ensuremath{\PZ\to\ell^{+}\ell^{-}}\xspace}
\newcommand{\wlnu}{\ensuremath{\PW\to\ell\PGn}\xspace}
\newcommand{\lqs}{\ensuremath{\mathrm{LQ}_{\mathrm{S}}}\xspace}
\newcommand{\lqv}{\ensuremath{\mathrm{LQ}_{\mathrm{V}}}\xspace}
\newcommand{\topnu}{\ensuremath{\PQt\PGn}\xspace}
\newcommand{\bnu}{\ensuremath{\PQb\PGn}\xspace}
\newcommand{\cnu}{\ensuremath{\PQc\PGn}\xspace}
\newcommand{\btau}{\ensuremath{\PQb\PGt}\xspace}
\newcommand{\gbl}{\ensuremath{g_{\PQb_{\mathrm{L}}}}\xspace}
\newcommand{\gtl}{\ensuremath{g_{\PQt_{\mathrm{L}}}}\xspace}

\cmsNoteHeader{SUS-18-001}
\title{Constraints on models of scalar and vector leptoquarks decaying to a quark and a neutrino at $\sqrt{s}=13\TeV$}

\date{\today}

\abstract{
The results of a previous search by the CMS Collaboration for squarks and gluinos are reinterpreted to constrain models of leptoquark (LQ) production. The search considers jets in association with a transverse momentum imbalance, using the \mttwo\ variable. The analysis uses proton-proton collision data at $\sqrt{s}=13\TeV$, recorded with the CMS detector at the LHC in 2016 and corresponding to an integrated luminosity of \fullLumi. Leptoquark pair production is considered with LQ decays to a neutrino and a top, bottom, or light quark.  This reinterpretation considers higher mass values than the original CMS search to constrain both scalar and vector LQs. Limits on the cross section for LQ pair production are derived at the 95\% confidence level depending on the LQ decay mode. A vector LQ decaying with a 50\% branching fraction to \topnu, and 50\% to \btau, has been proposed as part of an explanation of anomalous flavor physics results.  In such a model, using only the decays to \topnu, LQ masses below 1530\GeV are excluded assuming the Yang--Mills case with coupling $\kappa = 1$, or 1115\GeV in the minimal coupling case $\kappa = 0$, placing the most stringent constraint to date from pair production of vector LQs.
}

\hypersetup{
pdfauthor={CMS Collaboration},
pdftitle={Constraints on models of scalar and vector leptoquarks decaying to a quark and a neutrino at sqrt(s) = 13 TeV},
pdfsubject={CMS},
pdfkeywords={CMS, physics, leptoquark}}

\maketitle

\section{Introduction}

Leptoquarks (LQ) are hypothetical particles with quantum numbers of both quarks and leptons~\cite{Pati:1973uk}.
The spin of an LQ state is either 0 (scalar LQ, denoted \lqs) or 1 (vector LQ, denoted \lqv).
Leptoquarks appear in theories beyond the standard model (SM)
such as grand unified theories~\cite{Pati:1973uk, PatiSalam, GeorgiGlashow, Fritzsch:1974nn},
technicolor models~\cite{Dimopoulos:1979es,Dimopoulos:1979sp,Technicolor, Lane:1991qh},
compositeness scenarios~\cite{LightLeptoquarks,Gripaios:2009dq},
and R parity~\cite{Farrar:1978xj} violating supersymmetry (SUSY)~\cite{Ramond:1971gb,Golfand:1971iw,Neveu:1971rx,Volkov:1972jx,Wess:1973kz,Wess:1974tw,Fayet:1974pd,Nilles:1983ge,Barbier:2004ez}.

A growing collection of anomalies have been observed in flavor physics by the
BaBar~\cite{Lees:2012xj,Lees:2013uzd}, Belle~\cite{Matyja:2007kt,Bozek:2010xy,Huschle:2015rga,Hirose:2016wfn},
and LHCb~\cite{Aaij:2013qta,Aaij:2014ora,Aaij:2015yra,Aaij:2015oid,Aaij:2017vbb} Collaborations.
These have been explained as hints of lepton flavor universality violation in both charged- and neutral-current processes.
Leptoquarks have been suggested as an explanation of these
results~\cite{Tanaka:2012nw,Sakaki:2013bfa,Dorsner:2013tla,Gripaios:2014tna,Bauer:2015knc,Buttazzo:2017ixm,Dorsner:2018ynv}.
In particular, the best fit model of Refs.~\cite{Buttazzo:2017ixm,Dorsner:2018ynv} predicts an \lqv with a mass of $\mathcal{O}(\TeV)$ decaying with 50\% branching fraction
to either a top quark and a neutrino (\topnu) or a bottom quark and a tau lepton (\btau).
Such a state would therefore be visible at the CERN LHC.

\begin{figure*}[htbp]
\centering
\includegraphics[width=0.40\textwidth]{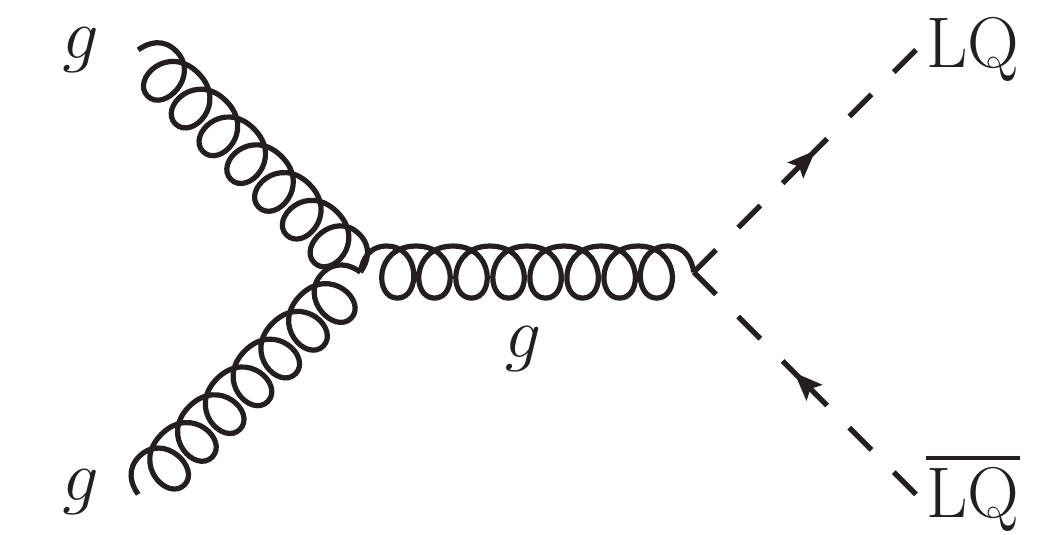}
\includegraphics[width=0.40\textwidth]{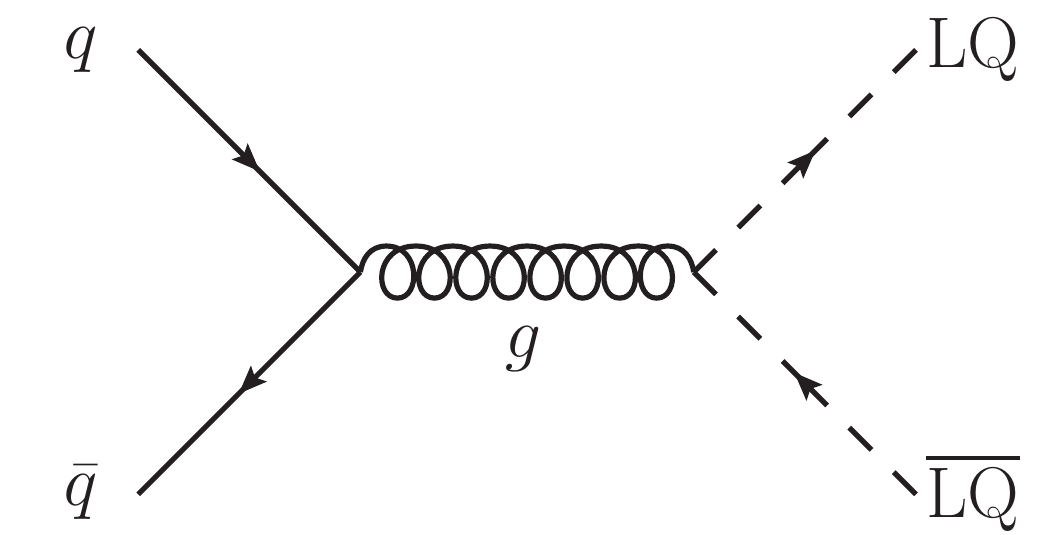}
\includegraphics[width=0.40\textwidth]{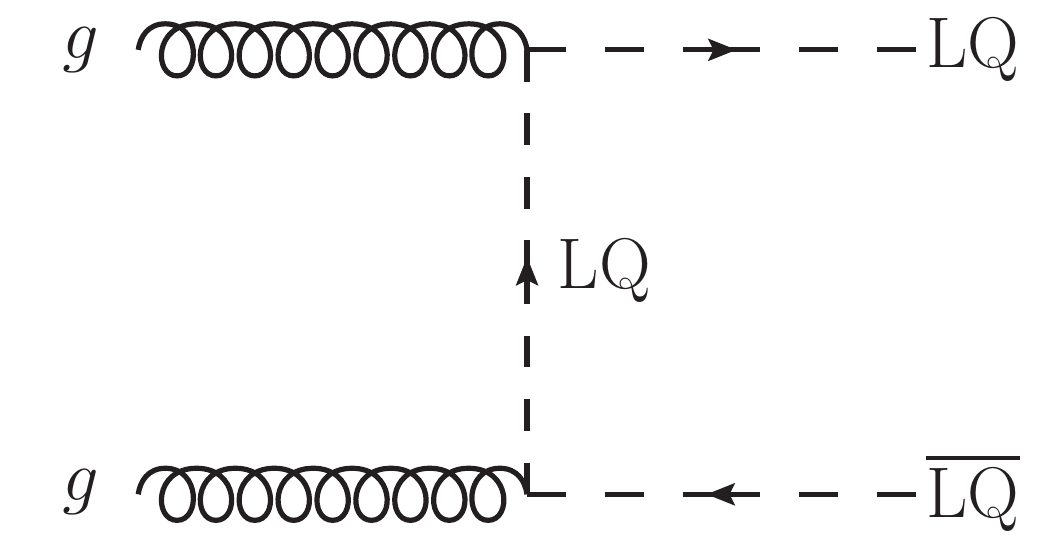}
\includegraphics[width=0.40\textwidth]{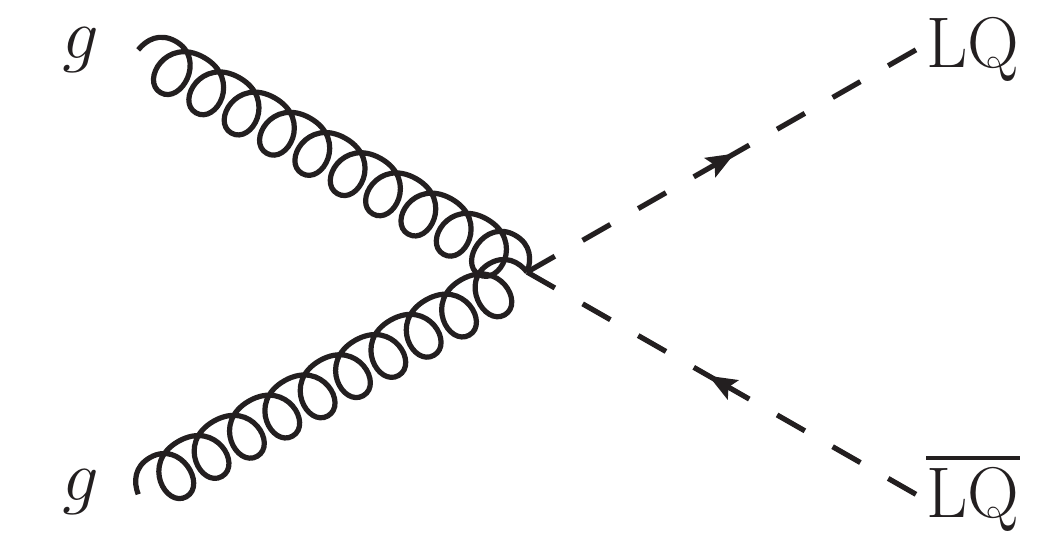}
\caption{Dominant LO diagrams for LQ pair production in proton-proton collisions.
}
\label{fig:diagrams}
\end{figure*}

At the LHC, LQ can be produced either in pairs or singly in association with a lepton.
In this paper, we focus on LQ pair production with both decaying to a neutrino
and a top, bottom, or light quark (any single one of up, down, strange, or charm).
The dominant leading-order (LO) diagrams for pair production at the LHC are shown in Fig.~\ref{fig:diagrams}.
The models for \lqs and \lqv pair production are taken from Ref.~\cite{Dorsner:2018ynv},
which provides a concrete implementation of the models from Ref.~\cite{Buttazzo:2017ixm}.
For \lqs, the pair production cross section depends only on the LQ mass.
For \lqv, there are additional constraints imposed by unitarity at high energy scales leading
to model dependent solutions and thus production cross sections.
In the model developed to explain the flavor physics anomalies~\cite{Dorsner:2018ynv},
the additional relevant parameter for the \lqv pair
production cross section is $\kappa$, a dimensionless coupling that is 1 in the Yang--Mills case
and 0 in the minimal coupling case, and we consider both values.
For $\kappa = 1$, the cross section for \lqv pair production is a factor of 5--20 times larger than
that of \lqs, depending on the LQ mass.
The other free parameters in the \lqv model are \gtl\ and \gbl,
the couplings of the \lqv to \topnu\ and \btau\ pairs respectively,
but they do not affect the cross section or kinematics for pair production.

The pair production of \lqs, each decaying to a quark and neutrino, results in the same final states and kinematics as those considered
in searches for squark pair production in R-parity conserving SUSY,
assuming that the squark decays directly to a quark and a massless neutralino~\cite{Diaz:2017lit}.
In both cases, the initial particles are scalars (\lqs or squark) produced strongly via quantum chromodynamics (QCD),
and the decay products are a quark and a nearly massless fermion (neutrino or neutralino).
In practice, the decay products in \lqv pair production are also found to have similar kinematics~\cite{Diaz:2017lit}.
Searches for squark pair production are therefore already optimized to search for LQ pair production.
Constraints on LQ production with decays to a quark and a neutrino have been placed using LHC data by the
ATLAS~\cite{Aad:2015caa} and CMS~\cite{Sirunyan:2018nkj,Khachatryan:2015vaa,Khachatryan:2014ura} Collaborations,
either by reinterpreting existing squark searches,
or considering mixed branching fraction scenarios with an LQ also decaying to a quark and a charged lepton.
Searches at the Tevatron covering the same signatures have been performed by the
CDF~\cite{Aaltonen:2009xp,Acosta:2005ge,Abulencia:2005ua} and D0~\cite{Abazov:2008at,Abazov:2011qj,Abazov:2008np} Collaborations.
Direct searches for single LQ production have also been performed at HERA
by the H1~\cite{Collaboration:2011qaa} and ZEUS~\cite{Abramowicz:2012tg} Collaborations,
placing constraints that are most stringent for an LQ with large coupling to an electron and a quark,
and large branching fraction for the decay to a quark and a neutrino.
Searches have also been performed by the ATLAS~\cite{ATLAS:2013oea}, CMS~\cite{Sirunyan:2017yrk},
CDF~\cite{Brigliadori:2008vf,Aaltonen:2007rb}, and D0~\cite{Abazov:2008jp} Collaborations
for an LQ decaying to the \btau\ channel as predicted in the model of Refs.~\cite{Buttazzo:2017ixm,Dorsner:2018ynv}.

The results from the CMS search for jets in association with a transverse momentum imbalance (\ptmiss)
using the \mttwo\ variable~\cite{MT2variable}, reported
in Ref.~\cite{mt2} and initially interpreted for squark and gluino production,
have recently been reinterpreted as part of a review of LQ searches to place
the strongest limits on the pair production of LQ decaying to a quark and a neutrino~\cite{Diaz:2017lit}.
However, for \lqv, the pair production cross sections are large enough that the mass range of
interest was not covered by the simulated samples used in Ref.~\cite{mt2}.
In particular, for an \lqv decaying to \topnu\ as predicted to explain the flavor physics anomalies,
the mass limit was derived from a flat extrapolation assuming that the cross section limit stayed the same at higher masses.
To improve upon these constraints, in this paper we present an extended interpretation of the search from Ref.~\cite{mt2},
where the selections, predictions, and uncertainties of the original analysis have not been changed.
Exploiting the similarity in final states between squark and LQ pair production,
we verify that the acceptance of our analysis is consistent within uncertainties for squark, \lqs, and \lqv pair production
for the same squark/LQ mass, assuming a neutralino mass of 1\GeV in the squark case.
We thus proceed to use simulated squark samples to place limits on both \lqs and \lqv production.
Using the full analysis information including all signal regions and correlations,
we extend the interpretations from Ref.~\cite{mt2} to higher mass values,
allowing us to improve the upper limits on LQ pair production cross sections in the \topnu\ decay channel
by as much as a factor of 2.8 over the flat extrapolation assumed in Ref.~\cite{Diaz:2017lit}.
With this approach, we derive the strongest coupling-independent constraints to date on the anomaly-inspired model
of Refs.~\cite{Buttazzo:2017ixm,Dorsner:2018ynv}.

\section{Analysis overview}
\label{sec:ana}

This study reinterprets the CMS search for jets and \ptmiss\ using the \mttwo\ variable.
The analysis is unchanged with respect to Ref.~\cite{mt2}, where a full description can be found, and is briefly summarized here.
The search uses proton-proton collision data at $\sqrt{s}=13\TeV$, recorded with the CMS detector in 2016,
and corresponding to an integrated luminosity of \fullLumi.
A description of the CMS detector, together with a definition of the coordinate system used and the relevant kinematic variables,
can be found in Ref.~\cite{Chatrchyan:2008zzk}.

Event reconstruction is based on the particle-flow (PF) algorithm~\cite{Sirunyan:2017ulk}.
Jets are clustered from PF candidates using the anti-\kt clustering algorithm~\cite{antikt} with a distance parameter of $\mathrm{R}=0.4$,
as implemented in the {\FASTJET} package~\cite{FastJet}, and are required to have pseudorapidity $\abs{\eta} < 2.4$.
Jets with transverse momentum $\pt > 20\GeV$ are identified as originating from \PQb\ quarks (``\PQb\ tagged'')
using the combined secondary vertex algorithm~\cite{Sirunyan:2017ezt},
and the number of \PQb-tagged jets is denoted \nbtags.
For all other quantities considered in the analysis, jets are required to satisfy $\pt > 30\GeV$.
The number of passing jets is denoted \njets, and the variable \HT\ is defined as the scalar sum of jet \pt.
The missing transverse momentum vector, \ptvecmiss, is defined as the negative vector sum of the momenta of all
reconstructed PF candidates originating from the primary vertex, projected onto the plane perpendicular to the proton beams.
Its magnitude is referred to as \ptmiss.

At the trigger level, events are selected by requiring large \HT, jet \pt, or \ptmiss.
The trigger selections have efficiency greater than 98\% for events with offline reconstructed values of $\ptmiss > 250\GeV$ or $\HT > 1000\GeV$.
The baseline selection requires $\njets \geq 1$, and events must pass either $\ptmiss > 30\GeV$ if they have $\HT > 1000\GeV$,
or $\ptmiss > 250\GeV$ if they have $250 < \HT < 1000\GeV$.
Further baseline requirements include that \ptvecmiss\ is not aligned in the azimuthal angle $\phi$ with any of the four leading jets in \pt,
that the negative vector sum of jet transverse momenta, \vMht, is consistent with \ptvecmiss,
and that no loosely identified charged leptons or isolated tracks are found in the event.
For events with $\njets \geq 2$, the variable \mttwo\ is computed from the jets and the \ptvecmiss
as described in Ref.~\cite{mt2}.
The \mttwo\ variable takes on small values for events where the momentum imbalance arises from jet mismeasurement,
typical of the QCD multijet background, and it yields larger values in events with genuine \ptvecmiss.
The baseline selection for events with $\njets \geq 2$ requires $\mttwo > 200\GeV$,
which is raised to $\mttwo > 400\GeV$ for events with $\HT > 1500\GeV$ to further reject multijet background.

Events with $\njets \geq 2$ passing the baseline selection are categorized according to four variables:
\HT, \mttwo, \njets, and \nbtags.
Events with $\njets = 1$ are categorized according to the jet \pt and the presence or absence of a \PQb-tagged jet.
The analysis spans a wide range of kinematics and jet multiplicities, containing 213 search bins in total,
to maintain sensitivity to a variety of new physics signatures.

The SM backgrounds to the search comprise three classes of processes: \zjets\ production with the decay \znunu,
\wjets\ or \ttjets\ production with the decay \wlnu\ where the charged lepton is outside the acceptance or not identified (``lost lepton''),
and QCD multijet production where \ptvecmiss\ arises from jet mismeasurement.
Each of these backgrounds is predicted primarily from data control regions:
\zjets\ from \zll\ events, \wjets\ and \ttjets\ from events containing an identified electron or muon,
and QCD multijets from events where at least one of the jets is aligned in $\phi$ with \ptvecmiss.

\begin{figure*}[htbp]
\centering
\includegraphics[width=0.46\textwidth]{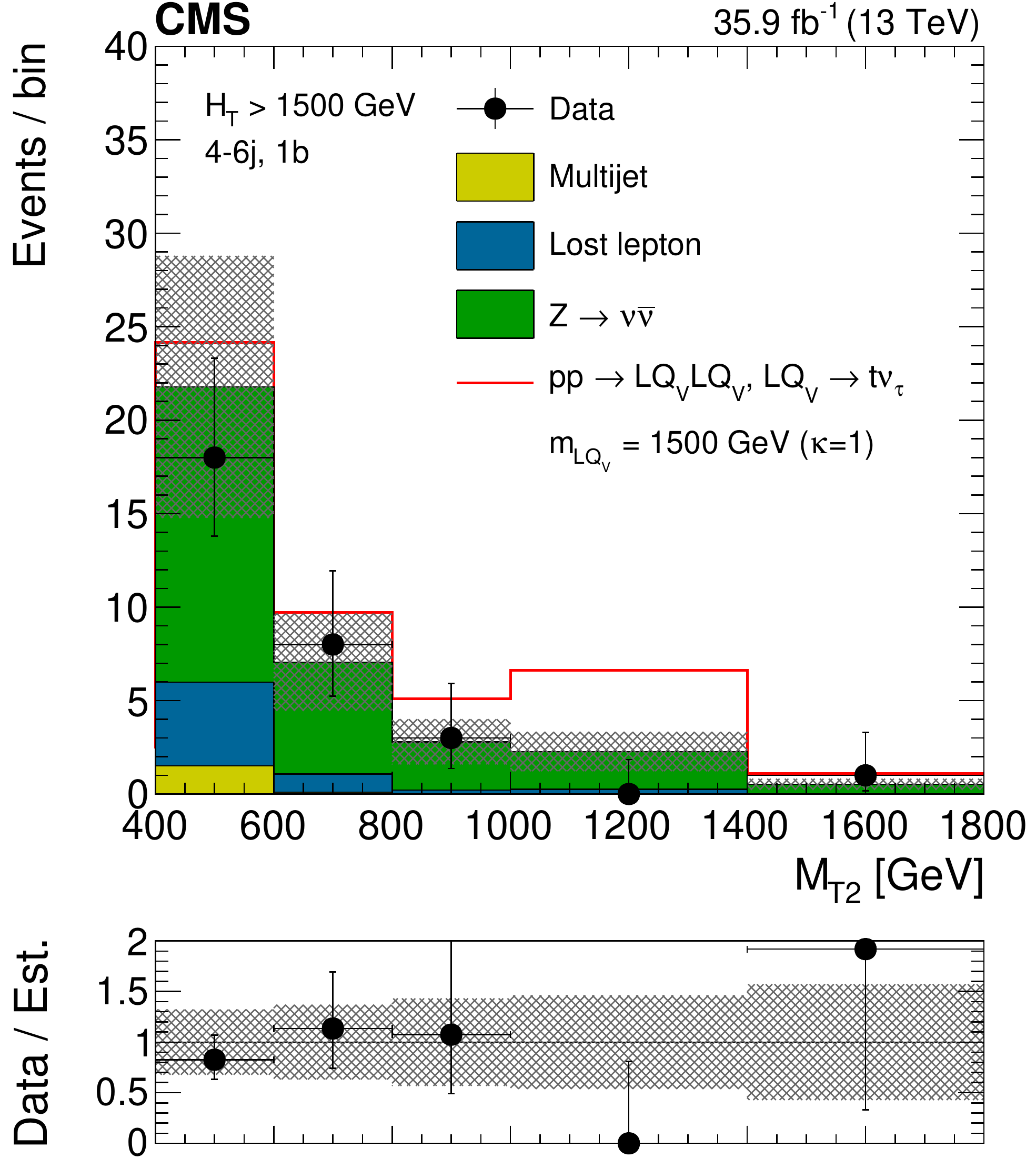}
\includegraphics[width=0.46\textwidth]{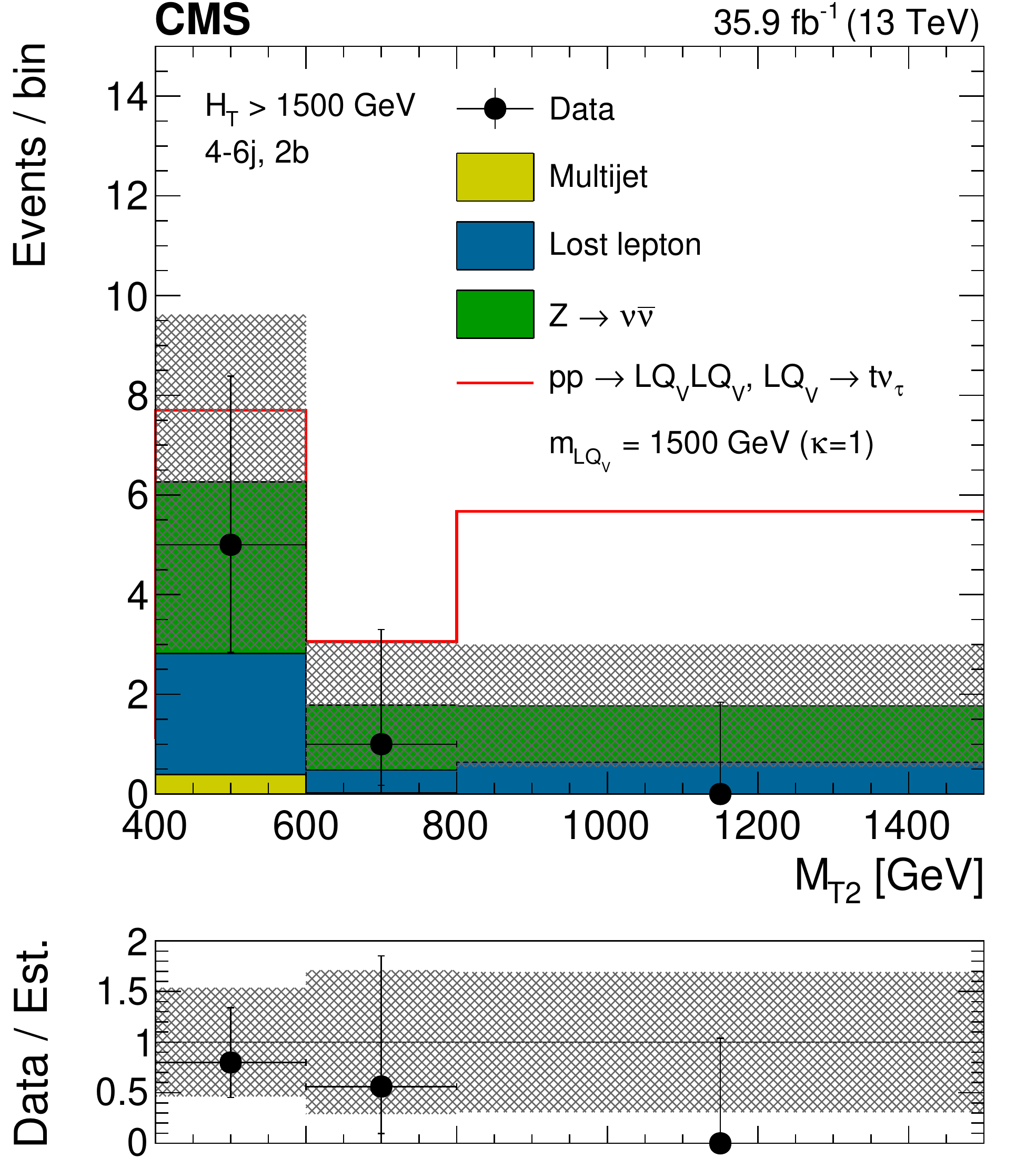}
\caption{Distributions of \mttwo showing data, the background predictions, and a hypothetical \lqv signal
with LQ mass of 1500\GeV decaying with 100\% branching fraction to $\topnu_{\tau}$.
The cross section used for the \lqv signal assumes $\kappa = 1$,
and the signal is stacked on top of the background predictions.
The black points show the observed data, with the statistical uncertainties represented by the vertical bars,
and the bin widths represented by the horizontal bars.
The rightmost bin in each plot also includes events with larger values of \mttwo.
The hatched band shows the uncertainty in the background prediction including both
statistical and systematic components.
The lower pane of each plot shows the ratio of observed data over predicted background.
The categories require $\HT > 1500\GeV$, $4 \leq \njets \leq 6$, and (left) $\nbtags = 1$ or
(right) $\nbtags = 2$.
}
\label{fig:mt2_dists}
\end{figure*}

Depending on the LQ mass and decay products, different search bins provide the greatest signal sensitivity.
Figure~\ref{fig:mt2_dists} shows the \mttwo\ distribution for data, the background predictions, and a hypothetical \lqv signal
in the two most sensitive search categories for an LQ of mass 1500\GeV decaying with 100\% branching fraction to \topnu.

Taking into account all of the analysis bins, no significant deviations from the SM prediction are observed.
Simultaneous maximum likelihood fits to data yields in all bins are performed,
and the results are interpreted as limits on the production cross sections of hypothetical scenarios of LQ pair production.

\section{Simulated samples}
\label{sec:mc}

Monte Carlo (MC) simulated samples are used to estimate the
background from some SM processes, to assess systematic uncertainties in prediction methods that rely on data,
and to calculate the selection efficiency for signal models.
The main background samples (\zjets, \wjets, and \ttjets), as well as signal samples,
are generated at LO precision in perturbative QCD with the \MGvATNLO\ v2.3.3 generator~\cite{Alwall:2014hca}.
Up to four, three, or two additional partons are considered in the matrix element calculations for the generation of the \vjets,
\ttjets, and signal samples, respectively.
The NNPDF3.0 LO~\cite{Ball:2014uwa} parton distribution functions (PDFs) are used in the event generation.
Parton showering and fragmentation are performed using the \PYTHIA\ v8.212~\cite{Sjostrand:2014zea} generator
and the CUETP8M1 tune~\cite{Khachatryan:2015pea}.
The potential double counting of the partons generated with \MGvATNLO\ and those with \PYTHIA\ is removed
using the MLM~\cite{Alwall:2007fs} matching scheme.
The samples used for the SM backgrounds are unchanged from Ref.~\cite{mt2},
and the details of the sample generation for other SM processes are described further there.

Additional proton-proton interactions in the same or nearby bunch crossings (pileup)
are generated with \PYTHIA\ and superimposed on the hard collisions.
The response of the CMS detector to SM background samples is simulated using a \GEANTfour-based model~\cite{Geant}, while that to new
physics signals is modeled using the CMS fast simulation package~\cite{fastsim}.
All simulated events are processed with the same chain of reconstruction programs as used for collision data.
Corrections are applied to simulated samples to account for differences between
measurements in data and the \GEANTfour simulation
of the trigger, \PQb\ tagging, and lepton selection efficiencies.
Additional differences arising from the fast simulation modeling of selection efficiencies, as well as
from the modeling of \ptmiss, are corrected in the fast simulation and included in the systematic uncertainties considered.

The generated signal samples used for this interpretation consist of
simplified models~\cite{bib-sms-1,bib-sms-2,bib-sms-3,bib-sms-4,Chatrchyan:2013sza}
of squark pair production, with the squark decaying to a quark of the same flavor and a neutralino with mass of 1\GeV.
Three samples are generated with different squark flavors:
``light'' squarks with an equal fraction of (\PSQu,\PSQd,\PSQs,\PSQc),
bottom squarks, and top squarks.
Squark masses up to 2300\GeV are generated, compared to Ref.~\cite{mt2}
where the generated samples extended to masses of
1800\GeV for light squarks, 1450\GeV for bottom squarks, and 1200\GeV for top squarks.
Below those mass values, the previous samples generated with the same configuration are used.

Samples of pair production of \lqs and \lqv are also generated for a limited number of LQ mass values,
to verify that the acceptance of the analysis at generator level is consistent with the squark samples used.
Samples of \lqs pair production are generated with the \PYTHIA\ v8.205 generator,
using the NNPDF2.3 LO~\cite{Ball:2012uwa} PDFs.
Samples of \lqv pair production are generated with the \MGvATNLO\ generator
at LO precision in perturbative QCD, including up to two additional partons in the matrix element calculations
and using the MLM matching scheme and NNPDF3.1 LO~\cite{Ball:2017uwa} PDFs.
The variables defined in Section~\ref{sec:ana} are computed at generator level, and
the kinematics of the generated squark samples are compared to those
of \lqs and \lqv pair production samples.
The acceptance of both the baseline analysis selection and the kinematic requirements for the most sensitive signal regions
is found to be consistent within statistical uncertainties of $\sim$3--10\% for the squark, \lqs, and \lqv samples.
The statistical uncertainty of the simulated signal samples is included
when using the squark samples to set limits on LQ pair production,
and based on this study no additional correction for, or systematic uncertainty in, the signal acceptance is applied.

To improve the modeling of the multiplicity of additional jets beyond those from the hard scatter process,
we weight the signal MC events based on the number of such jets, denoted \njetsisr\ for initial-state radiation (ISR).
The weighting factors are derived from a control region enriched in \ttbar\ events,
obtained by selecting events with exactly two leptons ($\Pe\Pe$, $\PGm\PGm$ or $\Pe\PGm$) and exactly two \PQb-tagged jets.
The factors are chosen to make the simulated jet multiplicity agree with data,
and they vary between 0.92 for $\njetsisr=1$ and 0.51 for $\njetsisr\geq 6$.
We take one half of the deviation from unity as the systematic uncertainty in these reweighting factors,
as an estimate of the differences between \ttbar and signal production.

The cross sections for \lqs or \lqv pair production are computed to next-to-leading-order (NLO) or LO precision
in perturbative QCD, following Ref.~\cite{Dorsner:2018ynv} and using the NNPDF2.3 NLO or LO PDF set, respectively.
In the \lqv model, we assume $\gtl = \gbl = 0.1$, and either $\kappa = 1$ or $\kappa = 0$,
as predicted to explain the flavor physics anomalies.
The uncertainties in cross section calculations arise from PDF variations and from the renormalization and factorization scale variations.
For PDF uncertainties, the NNPDF2.3 PDF set variations are used.
For scale uncertainties, renormalization and factorization scales are varied up and down by a factor of two with respect to the nominal values.
The theoretical uncertainties in the cross section are not included in the limit calculation but displayed separately
in Fig.~\ref{fig:lq_interp}.

\section{Interpretation}
\label{sec:interp}

The search results of Ref.~\cite{mt2} are interpreted to place cross section limits on LQ pair production as a function of the LQ mass.
A modified frequentist approach is used, employing the \CLs criterion and an asymptotic
formulation~\cite{Junk:1999kv,Read:2002hq,Cowan:2010js, ATL-PHYS-PUB-2011-011}.
The uncertainties in the signal acceptance and efficiency, and in the background predictions,
are incorporated as nuisance parameters. The observed data yields in control regions
are parameterized using gamma functions, while other nuisance parameters are
implemented using log-normal functions, whose widths reflect the size of the systematic uncertainty.

The following sources of uncertainty in the signal acceptance and efficiency are evaluated
and taken to be fully correlated across all analysis bins:
determination of the integrated luminosity~\cite{lumi2016}, trigger efficiency,
lepton identification and isolation efficiency,
lepton efficiency modeling in fast simulation, \PQb\ tagging efficiency, jet energy scale,
modeling of \ptmiss in fast simulation,
modeling of ISR, simulation of pileup,
and variations of the generator factorization and renormalization scales.
The statistical uncertainty of the simulated signal samples is taken to be uncorrelated in every bin.
The total uncertainty in the signal acceptance is typically around 5--25\% in the most sensitive analysis bins.
A detailed discussion of the uncertainties in the background prediction can be found in Ref.~\cite{mt2}.

\begin{figure*}[htbp]
\centering
\includegraphics[width=0.46\textwidth]{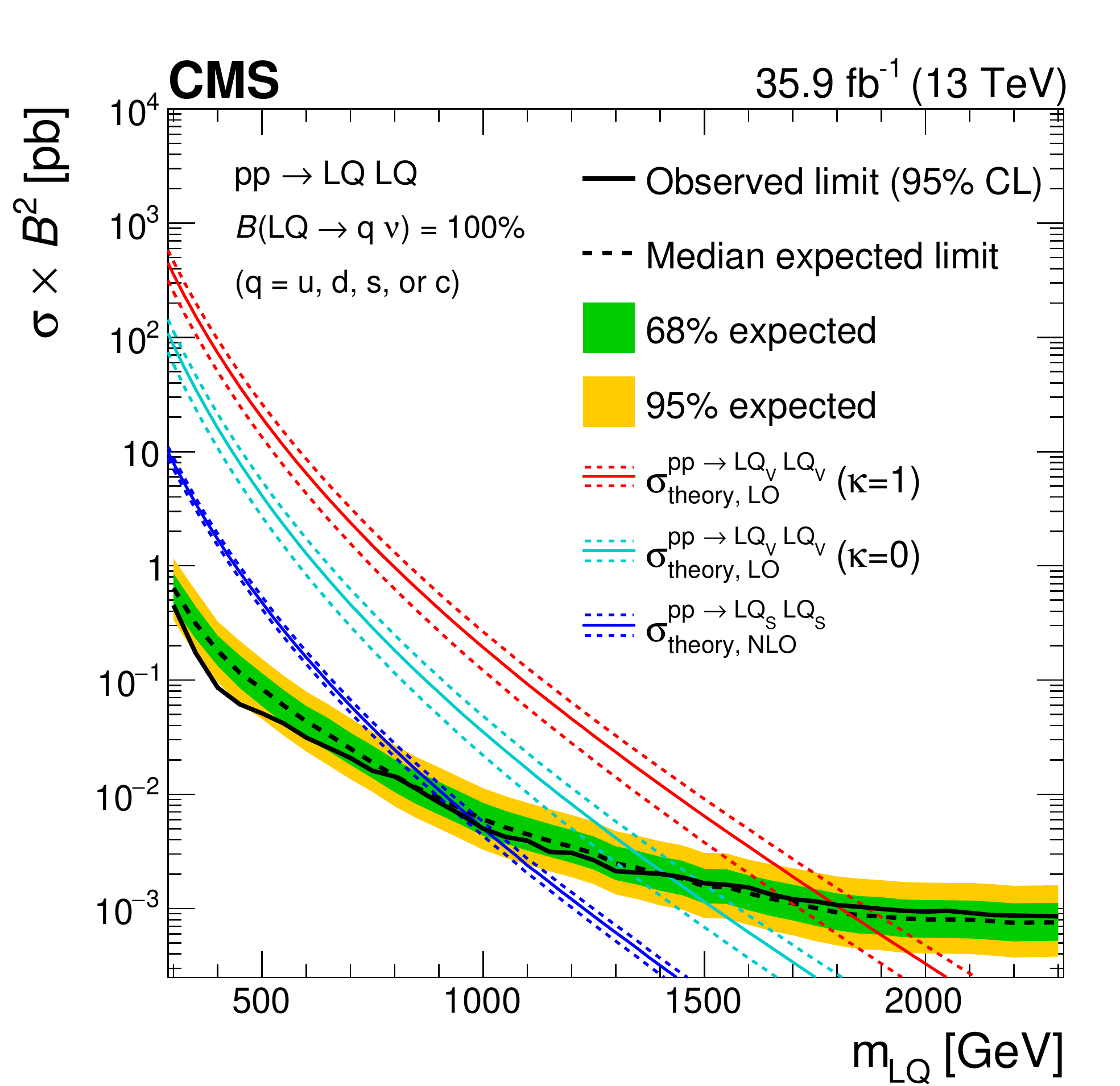}
\includegraphics[width=0.46\textwidth]{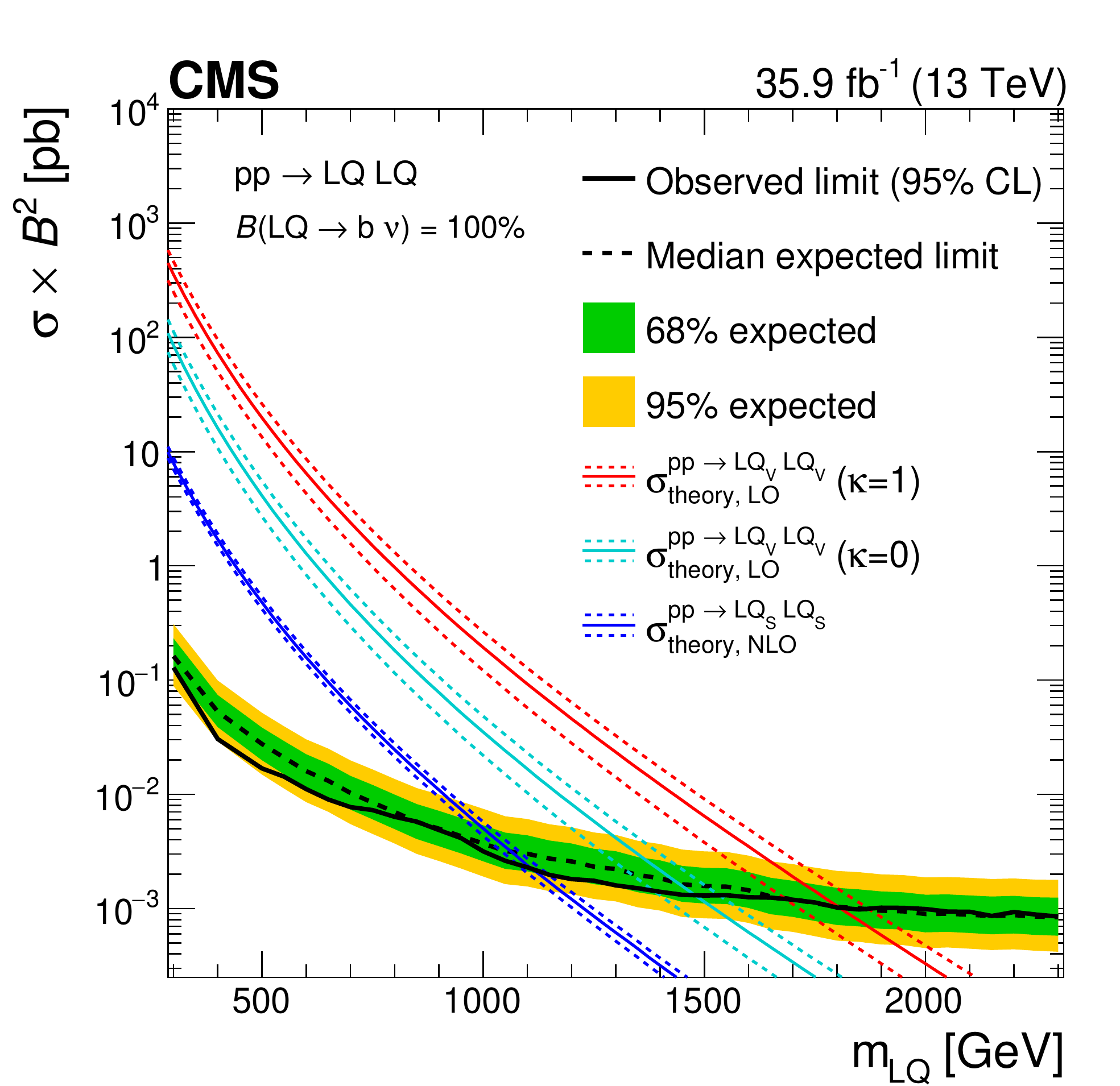}
\includegraphics[width=0.46\textwidth]{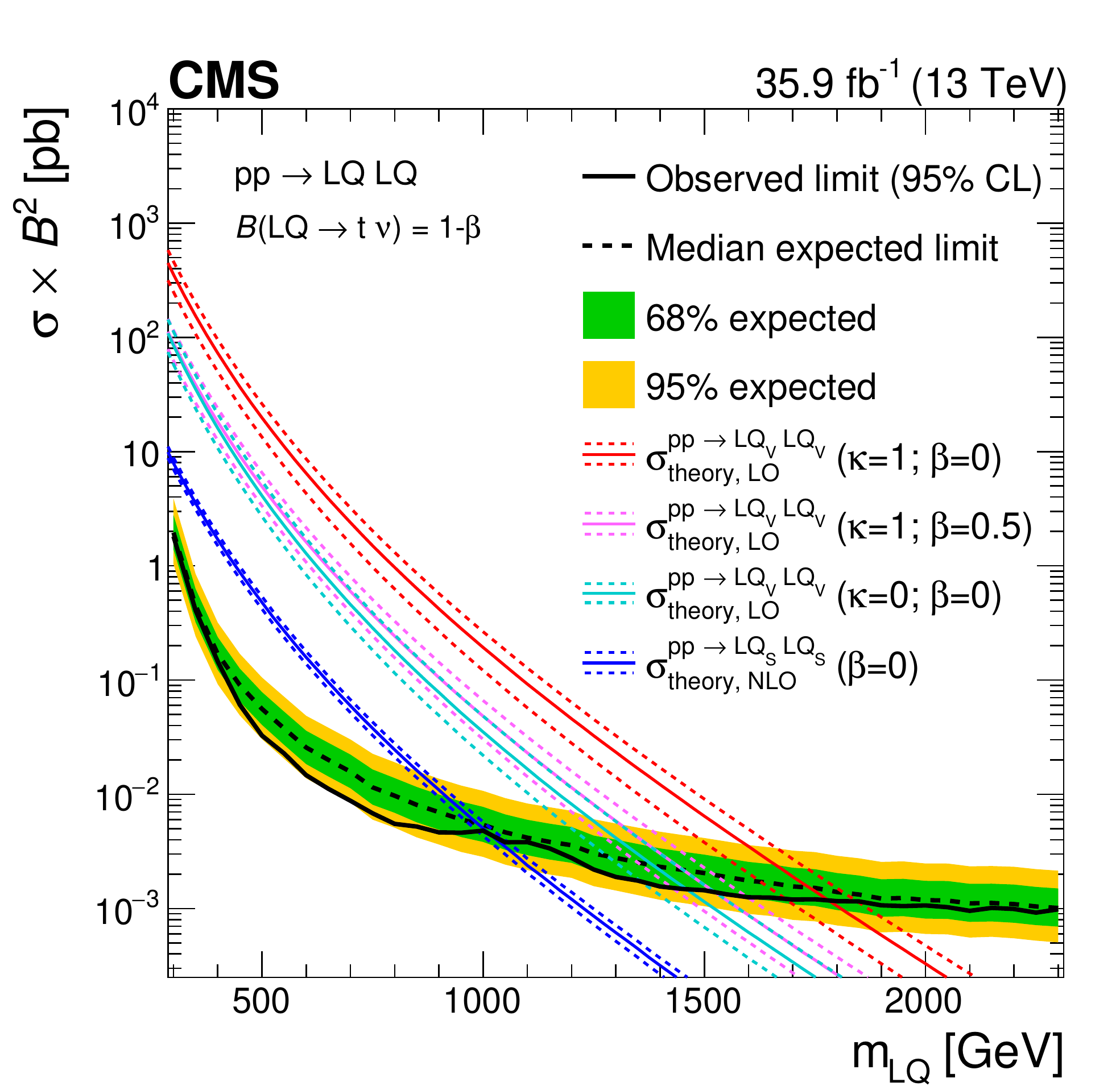}
\caption{The 95\% \CL upper limits on the production cross sections as a function of LQ mass
for LQ pair production decaying with 100\% branching fraction to a neutrino and (upper left) a light quark
(one of \PQu, \PQd, \PQs, or \PQc),
(upper right) a bottom quark, or (lower) a top quark.
The solid (dashed) black line represents the observed (median expected) exclusion.
The inner green (outer yellow) band indicates the region containing 68 (95\%)
of the distribution of limits expected under the background-only hypothesis.
The dark blue lines show the theoretical cross section for \lqs pair production with its uncertainty.
The red (light blue) lines show the same for \lqv\ pair production assuming $\kappa = 1$ (0).
(lower) Also shown in magenta is the product of the theoretical cross section and the square of the branching fraction ($B$),
for vector LQ pair production assuming $\kappa = 1$ and a 50\% branching fraction to $\topnu_{\tau}$, with the remaining 50\% to \btau.
}
\label{fig:lq_interp}
\end{figure*}

Exclusion limits at the 95\% confidence level (\CL) on the cross section of LQ pair production
are shown in Fig.~\ref{fig:lq_interp}.
In each case, we assume that there is only one LQ state with low enough mass to be produced at the LHC,
and that any other potential LQ states have masses too large to be produced.
We assume that the LQ decays with 100\% branching fraction to a neutrino and a single type of quark, as specified below.
In the simulated samples used to determine the signal acceptance, and for the cross sections displayed,
we consider only LQ pair production and not single LQ production.

We first consider LQ decays to a neutrino and a light quark, which can be any single one of the \PQu, \PQd, \PQs, or \PQc quarks.
As the analysis includes categorization in the number of \PQb-tagged jets,
and the probability for a \PQc quark to pass the \PQb\ tagging selection is larger than that of the \PQu, \PQd, and \PQs quarks,
we check whether the cross section limit obtained for an LQ decaying to \cnu\
differs significantly from an LQ decaying to a neutrino and one of the other light quarks.
The cross section limit differs by at most 10\%, resulting in a negligible impact on the mass limit,
and we therefore do not produce separate limit results for these cases.
The observed and expected limits on the LQ mass, and the corresponding cross sections for the excluded mass values,
are summarized in Table~\ref{tab:limits}.

\begin{table*}[htbp]
\centering
\topcaption{\label{tab:limits} Summary of the observed (expected) mass limits at the 95\% \CL,
and the cross sections $\sigma$ that correspond to the excluded mass values.
The columns show scalar or vector leptoquarks with the choice of $\kappa$,
while the rows show the LQ decay channel.
}
\cmsTable{
\begin{scotch}{c|cc|cc|cc}
 & \multicolumn{2}{c|}{\lqs} & \multicolumn{2}{c|}{\lqv, $\kappa = 1$} & \multicolumn{2}{c}{\lqv, $\kappa = 0$} \\
 & mass [\GeVns{}] & $\sigma$ [fb] & mass [\GeVns{}] & $\sigma$ [fb] & mass [\GeVns{}] & $\sigma$ [fb] \\
\hline
$\mathrm{LQ} \to \PQq\PGn$ & \multirow{2}{*}{980 (940)} & \multirow{2}{*}{5.9 (8.0)} & \multirow{2}{*}{1790 (1830)} & \multirow{2}{*}{1.1 (0.9)} & \multirow{2}{*}{1410 (1415)} & \multirow{2}{*}{2.0 (2.0)} \\
\PQq = \PQu, \PQd, \PQs, or \PQc & & & & & & \\
$\mathrm{LQ} \to \bnu$ & 1100 (1070) & 2.4 (3.0) & 1810 (1800) & 1.0 (1.1) & 1475 (1440) & 1.3 (1.7) \\
$\mathrm{LQ} \to \topnu$ & 1020 (980) & 4.3 (5.9) & 1780 (1740) & 1.2 (1.5) & 1460 (1385) & 1.5 (2.4) \\
$\mathrm{LQ} \to \begin{cases} \topnu\,(B=50\%) \\ \btau\,(B=50\%) \end{cases}$ & \NA & \NA & 1530 (1460) & 1.3 (2.1) & 1115 (1095) & 3.7 (4.2) \\
\end{scotch}
}
\end{table*}

The observed limit is more stringent than expected by up to two standard deviations in the LQ mass range of about 400--600\GeV
for a decay to a light or bottom quark and a neutrino,
and in the range of about 500--900\GeV for the \topnu\ decay channel.
The most sensitive analysis bins differ in each case, primarily in the \njets\ and \nbtags\ requirements.
The background estimates for these bins are derived from statistically independent control regions,
so the predictions and uncertainties are largely uncorrelated among these interpretations.

The model proposed in Refs.~\cite{Buttazzo:2017ixm,Dorsner:2018ynv} as an explanation of the flavor physics anomalies
predicts an \lqv with 50\% branching fraction to each of the \topnu\ and \btau\ channels.
As our analysis removes events with charged leptons, including hadronically decaying $\PGt$ leptons,
we only consider the 25\% of events where both LQ decay to \topnu\ to place constraints on this model.
We show the theoretical prediction for this branching fraction as a separate curve in Fig.~\ref{fig:lq_interp}~(lower)
assuming $\kappa = 1$.
For $\kappa = 1$, we find an observed (expected) limit on the \lqv mass of 1530 (1460)\GeV,
while for $\kappa = 0$ we obtain a limit of 1115 (1095)\GeV.
The mass limit in the $\kappa = 1$ case corresponds to a value of 1.3 (2.1)\unit{fb}
for the product of the LQ pair production cross section and the square of the branching fraction,
while for $\kappa = 0$, the value is 3.7 (4.2)\unit{fb}.

\section{Summary}

The CMS search for jets and missing transverse momentum using the \mttwo\ variable has been reinterpreted
to place limits on leptoquark (LQ) pair production, where the LQ decays with 100\% branching fraction to a quark and a neutrino.
The search uses proton-proton collision data at $\sqrt{s}=13\TeV$, recorded with the CMS detector in 2016
and corresponding to an integrated luminosity of \fullLumi.
Leptoquark decays to a neutrino and a top, bottom, or light quark are considered.
Compared to the original result, higher masses are considered to place exclusion limits on scalar LQs,
and on vector LQs assuming either the Yang--Mills ($\kappa = 1$) or minimal ($\kappa = 0$) coupling scenarios.
Assuming that there is only one LQ state within mass reach of the LHC
and that it decays to a light quark and a neutrino,
masses below 980, 1790, and 1410\GeV are excluded at the 95\% confidence level by the observed data
in the scalar, vector $\kappa = 1$, and vector $\kappa = 0$ scenarios.
For an LQ decaying to \bnu, masses below 1100, 1810, and 1475\GeV are excluded,
while for an LQ decaying to \topnu, masses below 1020, 1780, and 1460\GeV are excluded.
In the model of Refs.~\cite{Buttazzo:2017ixm,Dorsner:2018ynv}, a vector LQ with 50\% branching fraction
to \topnu, and 50\% to \btau, is predicted to explain anomalous flavor physics results.
Masses below 1530 (1115)\GeV are excluded for such a state assuming $\kappa = 1$ ($\kappa = 0$),
considering only the events with both LQ decaying to \topnu,
providing the strongest constraint to date in this model from pair production.

\begin{acknowledgments}
We congratulate our colleagues in the CERN accelerator departments for the excellent performance of the LHC and thank the technical and administrative staffs at CERN and at other CMS institutes for their contributions to the success of the CMS effort. In addition, we gratefully acknowledge the computing centers and personnel of the Worldwide LHC Computing Grid for delivering so effectively the computing infrastructure essential to our analyses. Finally, we acknowledge the enduring support for the construction and operation of the LHC and the CMS detector provided by the following funding agencies: BMWFW and FWF (Austria); FNRS and FWO (Belgium); CNPq, CAPES, FAPERJ, and FAPESP (Brazil); MES (Bulgaria); CERN; CAS, MoST, and NSFC (China); COLCIENCIAS (Colombia); MSES and CSF (Croatia); RPF (Cyprus); SENESCYT (Ecuador); MoER, ERC IUT, and ERDF (Estonia); Academy of Finland, MEC, and HIP (Finland); CEA and CNRS/IN2P3 (France); BMBF, DFG, and HGF (Germany); GSRT (Greece); NKFIA (Hungary); DAE and DST (India); IPM (Iran); SFI (Ireland); INFN (Italy); MSIP and NRF (Republic of Korea); LAS (Lithuania); MOE and UM (Malaysia); BUAP, CINVESTAV, CONACYT, LNS, SEP, and UASLP-FAI (Mexico); MBIE (New Zealand); PAEC (Pakistan); MSHE and NSC (Poland); FCT (Portugal); JINR (Dubna); MON, RosAtom, RAS and RFBR (Russia); MESTD (Serbia); SEIDI, CPAN, PCTI and FEDER (Spain); Swiss Funding Agencies (Switzerland); MST (Taipei); ThEPCenter, IPST, STAR, and NSTDA (Thailand); TUBITAK and TAEK (Turkey); NASU and SFFR (Ukraine); STFC (United Kingdom); DOE and NSF (USA).

\hyphenation{Rachada-pisek} Individuals have received support from the Marie-Curie program and the European Research Council and Horizon 2020 Grant, contract No. 675440 (European Union); the Leventis Foundation; the A. P. Sloan Foundation; the Alexander von Humboldt Foundation; the Belgian Federal Science Policy Office; the Fonds pour la Formation \`a la Recherche dans l'Industrie et dans l'Agriculture (FRIA-Belgium); the Agentschap voor Innovatie door Wetenschap en Technologie (IWT-Belgium); the F.R.S.-FNRS and FWO (Belgium) under the ``Excellence of Science - EOS" - be.h project n. 30820817; the Ministry of Education, Youth and Sports (MEYS) of the Czech Republic; the Lend\"ulet (``Momentum") Program and the J\'anos Bolyai Research Scholarship of the Hungarian Academy of Sciences, the New National Excellence Program \'UNKP, the NKFIA research grants 123842, 123959, 124845, 124850 and 125105 (Hungary); the Council of Science and Industrial Research, India; the HOMING PLUS program of the Foundation for Polish Science, cofinanced from European Union, Regional Development Fund, the Mobility Plus program of the Ministry of Science and Higher Education, the National Science Center (Poland), contracts Harmonia 2014/14/M/ST2/00428, Opus 2014/13/B/ST2/02543, 2014/15/B/ST2/03998, and 2015/19/B/ST2/02861, Sonata-bis 2012/07/E/ST2/01406; the National Priorities Research Program by Qatar National Research Fund; the Programa Estatal de Fomento de la Investigaci{\'o}n Cient{\'i}fica y T{\'e}cnica de Excelencia Mar\'{\i}a de Maeztu, grant MDM-2015-0509 and the Programa Severo Ochoa del Principado de Asturias; the Thalis and Aristeia programs cofinanced by EU-ESF and the Greek NSRF; the Rachadapisek Sompot Fund for Postdoctoral Fellowship, Chulalongkorn University and the Chulalongkorn Academic into Its 2nd Century Project Advancement Project (Thailand); the Welch Foundation, contract C-1845; and the Weston Havens Foundation (USA).
\end{acknowledgments}

\bibliography{auto_generated}
\cleardoublepage \appendix\section{The CMS Collaboration \label{app:collab}}\begin{sloppypar}\hyphenpenalty=5000\widowpenalty=500\clubpenalty=5000\vskip\cmsinstskip
\textbf{Yerevan Physics Institute, Yerevan, Armenia}\\*[0pt]
A.M.~Sirunyan, A.~Tumasyan
\vskip\cmsinstskip
\textbf{Institut f\"{u}r Hochenergiephysik, Wien, Austria}\\*[0pt]
W.~Adam, F.~Ambrogi, E.~Asilar, T.~Bergauer, J.~Brandstetter, M.~Dragicevic, J.~Er\"{o}, A.~Escalante~Del~Valle, M.~Flechl, R.~Fr\"{u}hwirth\cmsAuthorMark{1}, V.M.~Ghete, J.~Hrubec, M.~Jeitler\cmsAuthorMark{1}, N.~Krammer, I.~Kr\"{a}tschmer, D.~Liko, T.~Madlener, I.~Mikulec, N.~Rad, H.~Rohringer, J.~Schieck\cmsAuthorMark{1}, R.~Sch\"{o}fbeck, M.~Spanring, D.~Spitzbart, A.~Taurok, W.~Waltenberger, J.~Wittmann, C.-E.~Wulz\cmsAuthorMark{1}, M.~Zarucki
% ...
\vskip\cmsinstskip
% ...
1:  Also at Vienna University of Technology, Vienna, Austria\\
% ...
\end{sloppypar}
\end{document}